\documentclass[12pt]{article}
\usepackage{graphicx}
\usepackage{amsmath}

\input{tcilatex}

\begin{document}

\begin{titlepage}
\begin{flushright}
CAMS/00-03\\
\end{flushright}
\vspace{.6cm}
\begin{center}
\renewcommand{\thefootnote}{\fnsymbol{footnote}}
{\Large \bf Magnetic and Dyonic Black Holes In D=4 Gauged Supergravity}
\vfill
{\large \bf {A. H. Chamseddine \footnote{email: chams@aub.edu.lb} and
W.~A.~Sabra\footnote{email: ws00@aub.edu.lb}}}\\
\renewcommand{\thefootnote}{\arabic{footnote}}
\setcounter{footnote}{0}
\vfill
{\small
Center for Advanced Mathematical Sciences (CAMS)\\
and\\
Physics Department, American University of Beirut, Lebanon.}
\end{center}
\vfill
\begin{center}
{\bf Abstract}
\end{center}
Magnetic and Dyonic solutions are constructed for the theory of abelian gauged 
$N=2$ gauged four dimensional supergravity coupled to vector multiplets. 
The solutions found preserve 1/4 of the supersymmetry. 
\end{titlepage}
\vspace{1cm}

\section{Introduction}

There has been lots of interest in the study of solutions of gauged
supergravity theories in recent years \cite{bcs1, ck, klemm, bcs2,
duff,liu,sabra,chamsabra, klemmsabra}. Anti- de Sitter (AdS) black holes
solutions and their supersymmetric properties, in the context of $\ $gauged $%
N=2$, $D=4$ supergravity theory, were first considered in \cite{romans,
perry}. One of the motivations for the renewed interest is the fact that the
ground state of gauged supergravity theories is AdS space-time and
therefore, solutions of these theories may have implications for the
AdS/conformal field theory correspondence proposed in \cite{ads}. The
classical supergravity solution can provide some important information on
the dual gauge theory in the large $N$ (the rank of the gauge group). An
example of this is the Hawking-Page phase transition \cite{hawkpage, witten}%
. Also, the proposed AdS/CFT equivalence opens the possibility for a
microscopic understanding of the Bekenstein-Hawking entropy of
asymptotically anti-de~Sitter black holes \cite{strominger, brown}.
Moreover, AdS spaces are known to admit ``topological'' black holes with
some unusual geometric and physical properties (see for example \cite{top}). 
\newline

Of particular interest in this context are black objects in AdS space which
preserve some fraction of supersymmetry. On the CFT side, these supergravity
vacua could correspond to an expansion around non-zero vacuum expectation
values of certain operators.

Although lots is known about solutions in ungauged supergravity theories 
\cite{ungauged}, the situation is different for the gauged cases. Our main
purpose in this paper is to obtain supersymmetric solutions for a very large
class of supergravity theories coupled to vector multiplets. The theories we
will consider are four dimensional abelian gauged $N=2$ supergravity
theories coupled to vector multiplets. Electrically charged solutions of
these theories, asymptotic to ${\hbox{AdS}}_{4}$ space-time, were
constructed in \cite{sabra}. In this paper we will be mainly concerned with
the construction of magnetic as well as dyonic solutions.

Our work in this paper is organized as follows. Section two contains a brief
review of $D=4$, $N=2$ gauged supergravity where we collect formulae of
special geometry \cite{v} necessary for our discussion and for completeness
we present the electrically charged solutions of \cite{sabra}. In section
three, supersymmetric magnetic and dyonic solutions are constructed together
with their Killing spinors. Finally, our results are summarized and
discussed.\bigskip

\section{Gauged Supergravity and Special Geometry}

The construction of BPS solutions of the theory of ungauged and gauged $N=2$
supergravity coupled to vector supermultiplets, relies very much on the
structure of special geometry underlying these theories. The complex scalars 
$z^{A}$ of the vector supermultiplets of the $N=2$ supergravity theory are
coordinates which parametrize a special K\"{a}hler manifold. The Abelian
gauging is achieved by introducing a linear combination of the abelian
vector fields $A_{\mu }^{I}$ of the theory, $A_{\mu }={\kappa }_{I}A_{\mu
}^{I}$ with a coupling constant $g$, where ${\kappa }_{I}$ are constants.
The couplings of the fermi-fields to this vector field break supersymmetry
which in order to maintain one has to introduce gauge-invariant $g$%
-dependent terms. In a bosonic background, these additional terms produce a
scalar potential \cite{bigone} 
\begin{equation}
V=g^{2}\left( g^{A\bar{B}}{\kappa }_{I}{\kappa }_{J}f_{A}^{I}\bar{f_{\bar{B}%
}^{J}}-3{\kappa }_{I}{\kappa }_{J}\bar{L}^{I}L^{J}\right) .  \label{po}
\end{equation}

The meaning of the various quantities in (\ref{po}) is as follows. Roughly
one defines a special K\"{a}hler manifold in terms of a flat $(2n+2)-$
dimensional symplectic bundle over the K\"{a}hler-Hodge manifold, with the
covariantly holomorphic sections 
\begin{equation}
V=\left( 
\begin{array}{c}
L^{I} \\ 
M_{I}
\end{array}
\right) ,\text{ \ \ }I=0,\cdots ,n.\qquad \;D_{\bar{A}}V=(\partial _{\bar{A}%
}-{\frac{1}{2}}\partial _{\bar{A}}K)V=0,  \notag
\end{equation}
obeying the symplectic constraint 
\begin{equation}
i\langle V|\bar{V}\rangle =i(\bar{L}^{I}M_{I}-L^{I}\bar{M}_{I})=1.
\label{con}
\end{equation}
The scalar metric appearing in the potential can then be expressed as
follows 
\begin{equation*}
g_{A\bar{B}}=-i\langle U_{A}|\bar{U}_{\bar{B}}\rangle
\end{equation*}
\ \ 

where \ \ \ 

\begin{equation*}
U_{A}=D_{A}V=(\partial _{A}+{\frac{1}{2}}\partial _{A}K)V\ \ =\left( 
\begin{array}{c}
f_{A}^{I} \\ 
h_{AI}
\end{array}
\right) .
\end{equation*}

In general, one writes $M_{I}=\mathcal{N}_{IJ}L^{J}\ ,h_{AI}={\bar{\mathcal{N%
}}}_{IJ}f_{A}^{J}\ $ and the metric can then be expressed as $g_{A\bar{B}%
}=-2(\func{Im}\mathcal{N})_{IJ}f_{A}^{I}\bar{f}_{\bar{B}}^{J}.$ Moreover,
special geometry implies the following useful relation 
\begin{equation}
g^{A\bar{B}}f_{A}^{I}\bar{f}_{\bar{B}}^{J}=-{\frac{1}{2}}(\func{Im}\mathcal{N%
})^{IJ}-{\bar{L}}^{I}{L}^{J}.  \label{sg}
\end{equation}

The supersymmetry transformations for the gauginos and the gravitino in a
bosonic background 
\begin{align}
\delta \,\psi _{\mu }& =\left( \mathcal{D}_{\mu }+{\frac{i}{4}}T_{ab}\gamma
^{ab}\gamma _{\mu }-ig{\kappa }_{I}A_{\mu }^{I}+{\frac{i}{2}}{g}{\kappa }%
_{I}L^{I}\gamma _{\mu }\right) \epsilon ,  \notag \\
\delta \lambda ^{A}& =\left( i\,\gamma ^{\mu }\partial _{\mu }z^{A}+i%
\mathcal{G}_{\rho \sigma }^{A}\gamma ^{\rho }\gamma ^{\sigma }-gg^{A\bar{B}}{%
\kappa }_{I}f_{\bar{B}}^{I}\right) \epsilon .  \label{ssg}
\end{align}
where $\mathcal{D}_{\mu }$ is the covariant derivative, $\mathcal{D}_{\mu }$ 
$=(\partial _{\mu }-\frac{1}{4}\omega _{\mu }^{ab}\gamma _{ab}+\frac{i}{2}$ $%
Q_{\mu }),$ where $\omega _{\mu }^{ab}$ is the spin connection and $Q_{\mu }$
is the K\"{a}hler connection, which locally can be represented by 
\begin{equation}
Q=-{\frac{i}{2}}\left( \partial _{A}Kdz^{A}-\partial _{\bar{A}}Kd\bar{z}%
^{A}\right) .
\end{equation}
The quantities $T_{\mu \nu }$ and $\mathcal{G}_{\rho \sigma }^{A}$ are given
by

\begin{eqnarray}
T_{\mu \nu } &=&2i({\hbox{Im}}\mathcal{N}_{IJ})L^{I}F_{\mu \nu }^{J}  \notag
\\
\mathcal{G}_{\rho \nu }^{\,A} &=&-g^{A\bar{B}}\,\bar{f}_{\bar{B}%
}^{I}\,\left( {\hbox{Im}}\mathcal{N}_{IJ}\right) F_{\ \rho \nu }^{\,J},
\label{us}
\end{eqnarray}

Electrically charged BPS solutions for the $N=2$, $D=4$ supergravity with
vector multiplets were constructed in \cite{sabra} and are given by 
\begin{align}
& ds^{2}=-\left( e^{2U}+g^{2}r^{2}e^{-2U}\right) dt^{2}+{\frac{1}{\left(
e^{2U}+g^{2}r^{2}e^{-2U}\right) }}dr^{2}+e^{-2U}r^{2}(d\theta ^{2}+\sin
^{2}\theta d\phi ^{2}),  \notag \\
& e^{-2U}=Y^{I}H_{I},  \notag \\
& i(\mathcal{F}_{I}(Y)-\bar{\mathcal{F}}_{I}(\bar{Y})=H_{I},\qquad Y^{I}={%
\bar{Y}}^{I}  \notag \\
& A_{t}^{I}=e^{2U}Y^{I}.
\end{align}
where $H_{I}=\kappa _{I}+\frac{q_{I}}{r},$ and $q_{I}$ are the electric
charges. These electric BPS solutions break half of supersymmetry where the
Killing spinor $\epsilon $ satisfies 
\begin{equation}
\epsilon =(a\gamma _{0}+b\gamma _{1})\epsilon .  \label{cond}
\end{equation}
where 
\begin{align}
a& ={\frac{1}{\sqrt{1+g^{2}r^{2}e^{-4U}}}},  \notag \\
b& =-{i\frac{gre^{-2U}}{\sqrt{1+g^{2}r^{2}e^{-4U}}}},
\end{align}
The solution for the Killing spinor is given by 
\begin{equation}
\epsilon (r)={\frac{1}{2\sqrt{gr}}}e^{\frac{igt}{2}}e^{-{\frac{1}{2}}\gamma
_{0}\gamma _{1}\gamma _{2}\theta }e^{-{\frac{1}{2}}\gamma _{2}\gamma
_{3}\phi }e^{U+T}\Big(\sqrt{f-1}-i\gamma _{1}\sqrt{f+1}\big)(1-\gamma
_{0})\epsilon _{0}
\end{equation}
where $\epsilon _{0}$ is an arbitrary constant spinor and ${T}=\int^{r}{%
\frac{1}{2r^{\prime }}}(1-r^{\prime }\partial _{r^{\prime }}U(r^{\prime
}))dr^{\prime }$.

\section{Supersymmetric solutions}

In this section, we find magnetic and dyonic BPS solutions (with constant
scalars) of the theory of abelian gauged $N=2$ supergravity coupled to
vector supermultiplets. The solutions found preserve a quarter of
supersymmetry. We consider the following general form for the metric 
\begin{equation}
ds^{2}=-e^{2A}dt^{2}+e^{2B}dr^{2}+r^{2}(d\theta ^{2}+\sin ^{2}\theta d\phi
^{2}).  \label{ma}
\end{equation}
The vielbeins of this metric can be taken as 
\begin{align}
e_{t}^{0}& =e^{A},\quad e_{r}^{1}=e^{B},\quad e_{\theta }^{2}=r,\quad
e_{\phi }^{3}=r\sin \theta ,  \notag \\
e_{0}^{t}& =e^{-A},\quad e_{1}^{r}=e^{-B},\quad e_{2}^{\theta }={\frac{1}{r}}%
\ ,\quad e_{3}^{\phi }={\frac{1}{r\sin \theta }}.  \label{v}
\end{align}
and the spin connections for the above metric are 
\begin{align}
\omega _{t}^{01}& =A^{\prime }e^{A-B},  \notag \\
\omega _{\theta }^{12}& =-e^{-B},  \notag \\
\omega _{\phi }^{13}& =-e^{-B}\sin \theta ,  \notag \\
\omega _{\phi }^{23}& =-\cos \theta .  \label{sc}
\end{align}

\subsection{\protect\bigskip Magnetic Solutions}

First we concentrate on the purely magnetic BPS solutions. Therefore, we
take the vector potential to have only one non-vanishing component, i. e., 
\begin{equation}
A_{\phi }^{I}=q^{I}\cos \theta ,\text{ \ \ \ \ \ \ }F_{\theta \phi
}^{I}=-q^{I}\sin \theta ,  \label{gf}
\end{equation}
Using the above ansatz for the gauge fields and the spin connections in (\ref
{sc}), the supersymmetry transformations for the gravitino (\ref{ssg}) give

\begin{align}
\delta \psi _{t}& =\left[ \partial _{t}-\frac{1}{2}A^{\prime }e^{A-B}\gamma
_{0}\gamma _{1}+\frac{i}{2}e^{A}T_{23}\gamma _{2}\gamma _{3}\gamma _{0}+%
\frac{i}{2}ge^{A}\kappa _{I}L^{I}\gamma _{0}\right] \epsilon ,  \notag \\
\delta \psi _{\theta }& =\left[ \partial _{\theta }+\frac{1}{2}e^{-B}\gamma
_{1}\gamma _{2}+\frac{i}{2}T_{23}r\gamma _{3}+\frac{i}{2}gr\kappa
_{I}L^{I}\gamma _{2}\right] \epsilon ,  \notag \\
\delta \psi _{r}& =\left[ \partial _{r}+\frac{i}{2}T_{23}\gamma _{2}\gamma
_{3}\gamma _{1}e^{B}+\frac{i}{2}g(\kappa _{I}L^{I})\gamma _{1}e^{B}\right]
\epsilon ,  \notag \\
\delta \psi _{\phi }& ={\LARGE [}\partial _{\phi }+\frac{1}{2}\cos \theta
\gamma _{2}\gamma _{3}+\frac{1}{2}e^{-B}\sin \theta \gamma _{1}\gamma
_{3}+r\sin \theta (-T_{23}\gamma _{2}+\frac{i}{2}g\kappa _{I}L^{I}\gamma
_{3})-ig\kappa _{I}A_{\phi }^{I}{\LARGE ]}\epsilon .  \label{grt}
\end{align}
and for purely magnetic solutions we have 
\begin{equation*}
T_{23}=2i({\hbox{Im}}\mathcal{N}_{IJ})L^{I}F_{23}^{J}.
\end{equation*}
We will find supersymmetric configuration admitting Killing spinors which
satisfy the following conditions 
\begin{align}
\gamma _{1}\epsilon & =i\epsilon ,  \notag \\
\gamma _{2}\gamma _{3}\epsilon & =i\epsilon .  \label{sbc}
\end{align}
Because of the double projection on the spinor $\epsilon ,$ it is
independent of the angular variables $\theta $ and $\phi .$ With the
conditions (\ref{sbc}), one obtains from the vanishing of the gravitino
supersymmetry transformations in (\ref{grt}) the following equations 
\begin{align}
-\frac{1}{2}A^{\prime }e^{-B}+\frac{i}{2}T_{23}+\frac{1}{2}g(\kappa
_{I}L^{I})& =0,  \notag \\
-\frac{1}{2}e^{-B}-\frac{i}{2}T_{23}r+\frac{1}{2}g(\kappa _{I}L^{I})r& =0, 
\notag \\
\frac{1}{2}\cos \theta -g(\kappa _{I}A_{\phi }^{I})& =0.  \label{three}
\end{align}

and

\begin{align}
\partial _{t}\epsilon & =0,  \notag \\
\partial _{\theta }\epsilon & =0,  \notag \\
\left[ \partial _{r}-\frac{i}{2}T_{23}e^{B}-\frac{1}{2}g(\kappa
_{I}L^{I})\gamma _{1}e^{B}\right] \epsilon & =0,  \notag \\
\partial _{\phi }\epsilon & =0.  \label{ksm}
\end{align}
If one takes the ansatz \ 
\begin{equation*}
A=-B,
\end{equation*}
then from the first two equations in (\ref{three}) one obtains 
\begin{align}
\partial \left( e^{-B}\right) & =iT_{23}+g(\kappa _{I}L^{I}),  \notag \\
e^{-B}& =-iT_{23}r+gr(\kappa _{I}L^{I}).  \label{de}
\end{align}
An obvious solution to the above equations can be obtained if we set $%
(\kappa _{I}L^{I})$ to a constant say $\kappa _{I}L^{I}$ =1. Moreover, from\
the third relation in (\ref{three}), one gets the following ''quantization
relation'' 
\begin{equation*}
\kappa _{I}q^{I}=\frac{1}{2g}.
\end{equation*}
Therefore, as a solution for the holomorphic sections we take real $L^{I}$
and imaginary $M_{I},$ with 
\begin{equation}
L^{I}=2g\ q^{I}.  \label{ss}
\end{equation}
Using the symplectic constraint (\ref{con}), 
\begin{equation}
i(\bar{L}^{I}M_{I}-L^{I}\bar{M}_{I})=iL^{I}(M_{I}-\bar{M}%
_{I})=2iL^{I}M_{I}=1.
\end{equation}
Then this relation together with (\ref{ss}) implies that the magnetic
central charge $Z_{m}=M_{I}q^{I}=-\frac{i}{4g}$, and therefore we obtain 
\begin{equation*}
T_{23}=2i({\hbox{Im}}\mathcal{N}_{IJ})L^{I}F_{23}^{J}=-2\frac{M_{I}q^{I}}{%
r^{2}}=\frac{i}{2gr^{2}},
\end{equation*}
and from (\ref{de}) we obtain the following solution 
\begin{equation*}
e^{-B}=-iT_{23}r+gr(\kappa _{I}L^{I})=gr+\frac{1}{2gr}.
\end{equation*}

Finally one has to check for the vanishing of the our the gaugino
supersymmetry transformation for our solution. The gaugino transformation is
given by 
\begin{equation}
\delta \lambda ^{A}=\left( i\,\gamma ^{\mu }\partial _{\mu }z^{A}+i\mathcal{G%
}_{\rho \sigma }^{A}\gamma ^{\rho }\gamma ^{\sigma }-gg^{A\bar{B}}{\kappa }%
_{I}f_{\bar{B}}^{I}\right) \epsilon  \label{fin}
\end{equation}
where $\mathcal{G}_{\rho \nu }^{\,A}=-g^{A\bar{B}}\,\bar{f}_{\bar{B}%
}^{I}\,\left( {\hbox{Im}}\mathcal{N}_{IJ}\right) F_{\ \rho \nu }^{\,J}$, $%
g^{A\bar{B}}$ is the inverse K\"{a}hler metric and $\bar{f}_{\bar{B}%
}^{I}=(\partial _{\bar{B}}+{\frac{1}{2}}\partial _{\bar{B}}K)\bar{L}^{I}$.
To demonstrate the vanishing of the gaugino supersymmetry variations for the
choice of $\epsilon $, it is more convenient to multiply with $f_{A}^{I}$.
This gives using our solution and the relations following from special
geometry (\ref{sg}), 
\begin{equation*}
f_{A}^{I}\delta \lambda ^{\alpha A}=\left( i\gamma ^{\mu }\partial _{\mu
}z^{A}(\partial _{A}+{\frac{1}{2}}\,\partial _{A}K)L^{I}+{\frac{i}{2}}%
(F_{\mu \nu }^{\,I}-iL^{I}T_{\mu \nu })\gamma ^{\mu }\gamma ^{\nu }+{\frac{g%
}{2}}{\hbox{Im}}\mathcal{N}^{IJ}{\kappa }_{J}+g{\kappa }_{J}L^{J}L^{I}%
\right) \epsilon
\end{equation*}
The above transformation vanishes provided 
\begin{eqnarray}
(F_{23}^{\,I}-iL^{I}T_{23}) &=&0,  \notag \\
{\frac{g}{2}}{\hbox{Im}}\mathcal{N}^{IJ}{\kappa }_{J}+g{\kappa }%
_{J}L^{J}L^{I} &=&0.  \label{gve}
\end{eqnarray}
Using our solution it can be easily seen that the above equations are
satisfied.

In summary, we have obtained a BPS magnetic solution preserving a quarter of
supersymmetry for the theories of abelian gauged $\ N=2,$ $D=4$ supergravity
theories with vector multiplets. This solution is given by

\begin{eqnarray*}
ds^{2} &=&-(\frac{1}{2gr}+gr)^{2}dt^{2}+(\frac{1}{2gr}%
+gr)^{-2}dr^{2}+r^{2}(d\theta ^{2}+\sin ^{2}\theta d\phi ^{2}), \\
L^{I} &=&2g\ q^{I},\text{ \ \ \ }A_{\phi }^{I}=q^{I}\cos \theta .
\end{eqnarray*}

\vfill\eject

\subsection{Dyonic Solutions}

In this section we generalize the previous discussion to include electric
charges and thus obtaining dyonic solutions. From the outset we set $A=-B$
in the ansatz (\ref{ma}) as well as the condition $\kappa _{I}L^{I}=1$. \
The supersymmetric transformation for the gravitino in this case gives 
\begin{align}
\delta \psi _{t}& =\left[ \partial _{t}+\frac{1}{2}B^{\prime }e^{-2B}\gamma
_{0}\gamma _{1}+\frac{i}{2}T_{23}\gamma _{2}\gamma _{3}\gamma _{0}e^{-B}+%
\frac{i}{2}T_{01}\gamma _{1}e^{-B}-ig\kappa _{I}A_{t}^{I}+\frac{i}{2}%
ge^{-B}\gamma _{0}\right] \epsilon ,  \notag \\
\delta \psi _{\theta }& =\left[ \partial _{\theta }+\frac{1}{2}e^{-B}\gamma
_{1}\gamma _{2}+\frac{i}{2}T_{23}r\gamma _{3}-\frac{i}{2}T_{01}r\gamma
_{0}\gamma _{1}\gamma _{2}+\frac{i}{2}gr\gamma _{2}\right] \epsilon ,  \notag
\\
\delta \psi _{r}& =\left[ \partial _{r}+\frac{i}{2}T_{23}\gamma _{2}\gamma
_{3}\gamma _{1}e^{B}+\frac{i}{2}T_{01}\gamma _{0}e^{B}+\frac{i}{2}g\gamma
_{1}e^{B}\right] \epsilon ,  \notag \\
\delta \psi _{\phi }& ={\large [}\partial _{\phi }+\frac{1}{2}\cos \theta
\gamma _{2}\gamma _{3}+\frac{1}{2}e^{-B}\sin \theta \gamma _{1}\gamma _{3} 
\notag \\
& +r\sin \theta (\frac{i}{2}T_{23}\gamma _{2}\gamma _{3}-\frac{i}{2}%
T_{01}\gamma _{0}\gamma _{1})\gamma _{3}+\frac{i}{2}g\gamma _{3}-ig\kappa
_{I}A_{\phi }^{I}{\Large ]}\epsilon ,  \label{dt}
\end{align}
where 
\begin{align*}
T_{23}& =2i\func{Im}\mathcal{N}_{IJ}L^{I}F_{23}^{J}, \\
T_{01}& =2i\func{Im}\mathcal{N}_{IJ}L^{I}F_{01}^{J}.
\end{align*}

\bigskip The dyonic solutions can now be easily obtained by modifying one of
the supersymmetry breaking conditions (\ref{sbc}) and imposing the following
\ conditions on the Killing spinors 
\begin{align}
(ia\gamma _{0}+b\gamma _{1})\epsilon & =i\epsilon ,  \notag \\
\gamma _{2}\gamma _{3}\epsilon & =i\epsilon .  \label{msbc}
\end{align}
clearly the coefficients $a$ and $b$ must satisfy the condition $%
a^{2}+b^{2}=1.$

\bigskip Using the conditions (\ref{msbc}), then from the vanishing of the
transformations (\ref{dt}) we obtain the equations

\begin{align}
-\frac{1}{2b}e^{-B}-\frac{i}{2}T_{23}r+\frac{i}{2}T_{01}\frac{a}{b}r+\frac{1%
}{2}gr& =0,  \notag \\
\frac{a}{2b}e^{-B}-\frac{i}{2}T_{01}r\frac{1}{b}& =0,  \notag \\
\frac{1}{2b}B^{\prime }e^{-B}+\frac{i}{2}T_{23}-\frac{i}{2}T_{01}\frac{a}{b}+%
\frac{1}{2}g& =0,  \notag \\
-\frac{a}{2b}B^{\prime }e^{-2B}-g\kappa _{I}A_{t}^{I}+\frac{i}{2}T_{01}\frac{%
1}{b}e^{-B}& =0.  \label{gre}
\end{align}

together with

\begin{align}
\partial _{t}\epsilon & =0,  \notag \\
\partial _{\theta }\epsilon & =0,  \notag \\
\left[ \partial _{r}-\frac{1}{2}T_{23}\gamma _{1}e^{B}+\frac{i}{2}%
T_{01}\gamma _{0}e^{B}+\frac{i}{2}g\gamma _{1}e^{B}\right] \epsilon & =0, 
\notag \\
\partial _{\phi }\epsilon & =0.  \label{ksd}
\end{align}
As an ansatz for the gauge and scalar fields we take

\begin{equation*}
A_{t}^{I}=\frac{Z_{e}L^{I}}{r},\text{ \ \ \ }A_{\phi }^{I}=q^{I}\cos \theta ,%
\text{ \ \ \ \ }F_{01}^{I}=\frac{Z_{e}L^{I}}{r^{2}},\text{ \ \ \ \ \ \ }%
F_{23}^{I}=-\frac{q^{I}}{r^{2}},\text{ \ \ \ \ \ }L^{I}=2gq^{I},
\end{equation*}
where $Z_{e}=L^{I}Q_{I},$ is the electric central charge, $Q_{I}$ being the
electric charge. Therefore we obtain 
\begin{align*}
T_{23}& =2i\func{Im}\mathcal{N}_{IJ}L^{I}F_{23}^{J}=-2\frac{Z_{m}}{r^{2}}=%
\frac{i}{2gr^{2}}, \\
T_{01}& =2i\func{Im}\mathcal{N}_{IJ}L^{I}F_{01}^{J}=-i\frac{Z_{e}}{r^{2}}.
\end{align*}
From (\ref{gre}) we obtain the following solution for the metric and the
functions $a$ and $b,$

\begin{align}
e^{-2B}& =J_{1}^{2}+J_{2}^{2}.  \notag \\
J_{1}& =iT_{01}r=\kappa _{I}A_{t}^{I}=\frac{Z_{e}}{r}.  \notag \\
J_{2}& =gr-iT_{23}r=gr+2i\frac{Z_{m}}{r}=gr+\frac{1}{2gr}  \notag \\
a& =J_{1}e^{B},\ \ b=J_{2}e^{B}  \label{fs}
\end{align}

From the vanishing of the gaugino transformations in the presence of
electric charges, we get the conditions ($\ref{gve})$ together with the
newcondition

\begin{equation*}
(F_{01}^{\,I}-iL^{I}T_{01})=0.
\end{equation*}
All these conditions can be seen to be satisfied \bigskip for our ansatz.

Finally, we summarize dyonic solutions of $N=2$, $D=4$ gauged supergravity
with vector multiplets by

\begin{eqnarray*}
ds^{2} &=&-\left( (gr+\frac{1}{2gr})^{2}+\frac{Z_{e}^{2}}{r^{2}}\right)
dt^{2}+\left( (gr+\frac{1}{2gr})^{2}+\frac{Z_{e}^{2}}{r^{2}}\right)
^{-1}dr^{2}+r^{2}(d\theta ^{2}+\sin ^{2}\theta d\phi ^{2}), \\
L^{I} &=&2gq^{I},\text{ \ \ \ }A_{\phi }=q^{I}\cos \theta ,\text{ \ \ \ }%
Z_{e}=L^{I}Q_{I}.
\end{eqnarray*}

\subsection{\protect\bigskip Solutions With Hyperbolic And Flat Transverse
Spaces}

Clearly the spherical solutions we found represent naked singularities.
However, one can obtain extremal purely magnetic solutions with event
horizons if the two sphere is replaced with the quotients of the hyperbolic
two-space $H^{2}$. In this case, it can be shown that one obtains the
following solitonic dyonic solution

\begin{eqnarray*}
ds^{2} &=&-\left( gr-\frac{1}{2gr}\right) ^{2}dt^{2}+\left( gr-\frac{1}{2gr}%
\right) ^{-2}dr^{2}+r^{2}(d\theta ^{2}+\sinh ^{2}\theta d\phi ^{2}). \\
A_{\phi }^{I} &=&q^{I}\cosh \theta ,\quad L^{I}=2gq^{I}.
\end{eqnarray*}
As the Killing spinors do not depend on the coordinates $\theta ,\phi $ of
the transverse hyperbolic space, one could also compactify the $H^{2}$ to a
Riemann surface $\mathcal{S}_{n}$ of genus $n$, and the resulting solution
would still preserve one quarter of supersymmetry.\newline
Whereas the spherical magnetic solution contains a naked singularity, the
hyperbolic magnetic black hole solution is a genuine black hole solution
which has an event horizon at $r=r_{+}=1/(g\sqrt{2})$. In the near horizon
region, the metric reduces to the product manifold $AdS_{2}\times H^{2}$ and
thus supersymmetry is enhanced \cite{kz}.

One can also consider ~flat transverse space with vanishing gauge fields.
Clearly one finds the solution which is locally $AdS_{4}.$ However, one may
wish to compactify the $(\theta ,\phi )$ sector to a cylinder or a torus,
considering thus a quotient space of $AdS_{4}$. This identification results
in $AdS_{4}$ quotient space preserving half of the supersymmetries.

\subsection{\protect\bigskip Killing spinors}

We now derive an expression for the Killing spinors of our solutions. From (%
\ref{ksd}), one obtain the following equations for the Killing spinors 
\begin{align}
\partial _{t}\epsilon & =0,  \notag \\
\partial _{\theta }\epsilon & =0,  \notag \\
\left( \partial _{r}+\frac{1}{2r}+ig\gamma _{1}e^{B}\right) \epsilon & =0, 
\notag \\
\partial _{\phi }\epsilon & =0.
\end{align}

\bigskip Using the method of \cite{romans}, one obtain the following
solution for the Killing spinor 
\begin{equation*}
\epsilon (r)=\left( \sqrt{e^{-B}+gr+\frac{1}{2gr}}-\gamma _{0}\sqrt{%
e^{-B}-gr-\frac{1}{2gr}}\right) P(-i\gamma _{1})P(-i\gamma _{23})\epsilon
_{0}
\end{equation*}
where $\epsilon _{0}$ is a constant spinor and where we have used the
notation $P(\Gamma )=\frac{1}{2}(1+\Gamma ),$ where $\Gamma $ is an operator
satisfying $\Gamma ^{2}=1.$

\bigskip For the purely magnetic solution, i. e, $Q_{I}^{{}}=0,$ the Killing
spinor reduces to the simple form 
\begin{equation*}
\epsilon (r)=\sqrt{gr+\frac{1}{2gr}}P(-i\gamma _{1})P(-i\gamma
_{23})\epsilon _{0}
\end{equation*}

It must be mentioned as was observed in \cite{romans} that for the purely
magnetic solution , $e^{-B}$ and $\epsilon (r)$ are invariant under

\begin{equation*}
r\longrightarrow \frac{1}{2g^{2}r}
\end{equation*}
and under such a transformation, the spatial component of the metric is
conformally rescaled. Clearly the Killing spinors for the hyperbolic case
are similar and can be obtained by replacing $gr+\frac{1}{2gr}$ by $gr-\frac{%
1}{2gr}$ everywhere\cite{ck}.

\section{Discussion}

In summary, we have obtained magnetic and dyonic BPS solutions of the theory
of abelian gauged $N=2$ supergravity coupled to a number of vector
supemultiplets. These solutions preserve a quarter of supersymmetry. Due to
the quantization of the magnetic central charge for these solutions, the
metric of the magnetic solutions takes a universal form for any choice of
the prepotential and for any number of vector multiplets (charges). The
dyonic solutions depend also on the electric central charge. We note that
our solution is expressed in terms of the holomorphic sections and the
central charges of the theory and therefore independent of the existence of
a holomorphic prepotential. A subclass of solutions of $N=2$ supergravity
(for a particular choice of prepotential) are actually also solutions of
supergravity theories with more supersymmetry ,( i.e. $N=4$ or $N=8$
supersymmetries). The spherically symmetric solutions has the common feature
of representing naked singularities. However, it is known from the work of 
\cite{perry} that the spherical Kerr-Newman-AdS solution is both
supersymmetric and extreme. This means that one can obtain supersymmetric
extremal black holes in an AdS background if the black holes are rotating.
Thus the situation in anti-de Sitter background seems to be the opposite of
that of the asymptotically flat.

Whereas the spherical BPS magnetic black holes contain a naked singularity,
the hyperbolic black holes have an event horizon. Supersymmetric higher
genus solution black holes with magnetic charges preserve the same amount of
supersymmetry which is enhanced near the horizon as has been demonstrated in 
\cite{kz} for the pure supergravity theory without vector multiplets. Here
we found that the situation remains the same in the presence of vector
multiplets. In summary, in order to get genuine black holes for the theories
we have considered in this paper, one should allow for rotations as well as
nonspherical symmetry. This will be reported on in a future publication.

\end{document}